\documentclass[aps,prb,twocolumn,showpacs]{revtex4}
\usepackage{epsfig}
\usepackage{hyperref}
\usepackage{longtable}
\begin{document}
\def\il{I_{low}} 
\def\iu{I_{up}} 
\def\eeq{\end{equation}}
\def\ie{i.e.}  
\def\etal{{\it et al. }}  
\def\prb{Phys. Rev. {\bf B}}
\def\pra{Phys. Rev. {\bf A}} 
\def\prl{Phys. Rev. Lett. }
\def\pla{Phys. Lett. A } 
\def\pb{Physica B}
\def\ajp{Am. J. Phys. }  
\def\mpl{Mod. Phys. Lett. {\bf B}} 
\def\ijmp{Int. J. Mod. Phys. {\bf B}} 
\def\ijp{Ind. J. Phys. }
\def\ijpap{Ind. J. Pure Appl. Phys. }
\def\ibmjrd{IBM J. Res. Dev. }
\def\pjp{Pramana J. Phys.}

\title{Study of quantum current enhancement, eigenenergy spectra and
  magnetic moments in a multiply connected system at equilibrium}

\author{Colin Benjamin} 
\email{colin@iopb.res.in}
\author{A. M. Jayannavar}
\email{jayan@iopb.res.in}
\affiliation{Institute of Physics, Sachivalaya Marg, Bhubaneswar 751 005,
  Orissa, India}

\date{\today}

\begin{abstract}
   A multiply connected system in both its open and closed form
   variations but in equilibrium is studied using quantum waveguide
   theory. The system exhibits remarkable features, in its open form
   variation we see current enhancement, hitherto seen only in
   non-equilibrium systems in absence of magnetic flux. In its closed
   form analog parity effects are broken. Further we analyse the
   global and local current densities of our system and also show that
   the orbital magnetic response of the system calculated from the
   current densities (and inherently linked to the topological
   configuration) is qualitatively not same as that calculated from
   the eigenenergy spectra.
\end{abstract}

\pacs{73.23.-b, 05.60.Gg, 72.10.Bg, 72.25.-b }
\maketitle

  Mesoscopic systems are those whose dimensions lie between the
  macroscopic and the atomic scale.  Typically the system sizes are in
  the nano-range. Thus, mesoscopic physics addresses fundamental
  problems which occur when a macroscopic object is miniaturized. Over
  the past decade, research into the electrical transport properties
  of mesoscopic systems has provided insight into fundamental
  questions in physics especially the role of quantum fluctuations,
  quantum mechanics of interacting electrons, the quantum classical
  crossover, and the gradual buildup of macroscopic classical
  behavior.  In these systems electron retains its quantum coherence
  over the entire sample and thus classical laws which hold in the
  macroscopic regime breakdown\cite{imry,datta,psd,webb}. Some of them
  being breakdown of classical Ohm's law, sample specific resistance
  and its fluctuations, quantization of point contact conductance,
  breakdown of Onsager's reciprocity relations, etc.
  
  If the system dimensions are less than the electron phase relaxation
  length then electron retains its phase coherence and scattering in
  the sample is only elastic, indeed electronic motion in such systems
  is not unlike light propagation in waveguide structures and
  Fabry-Perot interferometers. The study of open and closed mesoscopic
  rings has given rise to some surprising findings. In closed
  mesoscopic rings pierced by a magnetic flux persistent currents have
  been predicted\cite{bil} and also experimentally
  seen\cite{levy,chandra,maily}. The magnetic flux destroys the
  time-reversal symmetry and as a consequence persistent current flows
  in a ring.  These currents are periodic in magnetic flux, with a
  period $\Phi_0$, $\Phi_0$ being the elementary flux quanta
  ($\Phi_0=\frac{h c}{e}$).  At zero temperature in a ballistic ring
  of circumference $l$ the amplitude of persistent current is given by
  $\frac{e v_f}{l}$, where $v_f$ is the Fermi velocity. For, spinless
  electrons, the persistent current can be either diamagnetic or
  paramagnetic depending upon whether the total number of electrons
  present in the isolated ring is odd or even respectively. This
  behavior of the persistent current is also known as the parity
  effect. In open rings pierced by the same magnetic flux, normal
  state Aharonov-Bohm (AB) oscillations have been observed apart from
  this currently these AB oscillations in open rings are at the heart
  of a large number of experiments, especially some wherein the
  quantum phase shift is purported to have been measured by inserting
  a quantum dot in one of the arms of the ring\cite{aharony} and
  measurement of which path detection which is crucial in
  understanding the nature of dephasing via the role of quantum
  entanglement.
  
  Another, purely quantum mechanical phenomena in such mesoscopic
  rings although in the absence of flux is that of current enhancement
  or magnification\cite{deo_cm,pareek_cm,colin_cm}. Current
  enhancement can be defined as follows- In a metallic loop connected
  to two reservoirs at chemical potentials $\mu_{1}$ and $\mu_{2}$
  (with $\mu_{1}>\mu_{2}$) by means of two ideal leads (as in inset of
  figure~1), transport current $I$ flows through the system.  This
  transport current divides into $I_{1}$ and $I_2$ in the upper and
  lower arms of the ring.  In classical case both $I_{1}$ and $I_{2}$
  are positive and flow in the same direction as the input transport
  current. In quantum mechanics, however for particular values of
  Fermi energy intervals $I_{1}$ or $I_{2}$ can become much larger
  than I, this implies to obey Kirchoff's law the current in the other
  loop must be negative. This property that current in one of the arms
  is larger than the transport current is referred to as current
  enhancement effect. In this situation, we interpret the negative
  current flowing in one arm of the ring as a circulating current that
  flows continually in the loop. When the negative current flows in
  the upper arm the circulating current direction is taken to be
  anti-clockwise (or negative) and when it flows in the lower arm the
  circulating current direction is taken to be clockwise (or positive)
  \cite{physicapareek}.  Studies on current magnification effect in
  mesoscopic open rings have been extended to thermal
  currents\cite{mosk} and to spin currents in the presence of
  Aharonov-Casher flux\cite{choi}.  Recently this effect has been
  studied in presence of spin-flip scattering which causes dephasing
  of electronic motion\cite{colin_cm,joshi}.

  The current enhancement effect leads to an enhanced magnetic
  response (orbital magnetic moment) of a loop carrying current in the
  absence of magnetic flux\cite{psd}. It is to be noted that these
  circulating currents arise in the absence of magnetic flux, however,
  in presence of transport currents (i.e., in a non-equilibrium
  system).  In the present work our thrust is whether we can observe
  the aforesaid current enhancement effect and the resulting
  circulating currents in equilibrium. For this we consider the system
  as depicted in figure~1.  The static localised flux piercing the
  loop is necessary to break the time reversal symmetry and induce a
  persistent current in the system.  The geometry we consider is a
  one-dimensional ring with an attached bubble and a lead connected to
  a reservoir at chemical potential $\mu$, for simplicity we have
  ignored the electron-electron interaction. The reservoir acts as an
  inelastic scatterer and as a source of energy dissipation. All the
  scattering processes in the leads including the loop are assumed to
  be elastic. Hence there is a complete spatial separation between the
  elastic and inelastic processes. The loops J1J2aJ3J1 and J1J2bJ3J1
  enclose the localised flux $\Phi$. However, the bubble J2aJ3bJ2 does
  not enclose the flux $\Phi$. We have considered this special
  topology in order to verify the existence of circulating currents at
  equilibrium.  We show that circulating currents (due to current
  enhancement) arise in a bubble which does not enclose a magnetic
  flux. We would like to mention here that the current enhancement
  effect and the associated circulating currents arise even when the
  magnetic field extends over the entire sample.  However, for this
  the treatment is involved as one has to study separately persistent
  as well as circulating currents in the bubble as they have different
  symmetry properties. This has been studied in a simple loop in the
  presence of both transport currents and magnetic
  flux\cite{physicapareek}. Just for the sake of simplicity and to
  show the existence of current enhancement in equilibrium we have
  taken a system in which bubble does not enclose a magnetic flux,
  which may not be an ideal system. However, it clarifies our
  contention.

\begin{figure}{h}
\protect\centerline{\epsfxsize=3.0in \epsfbox{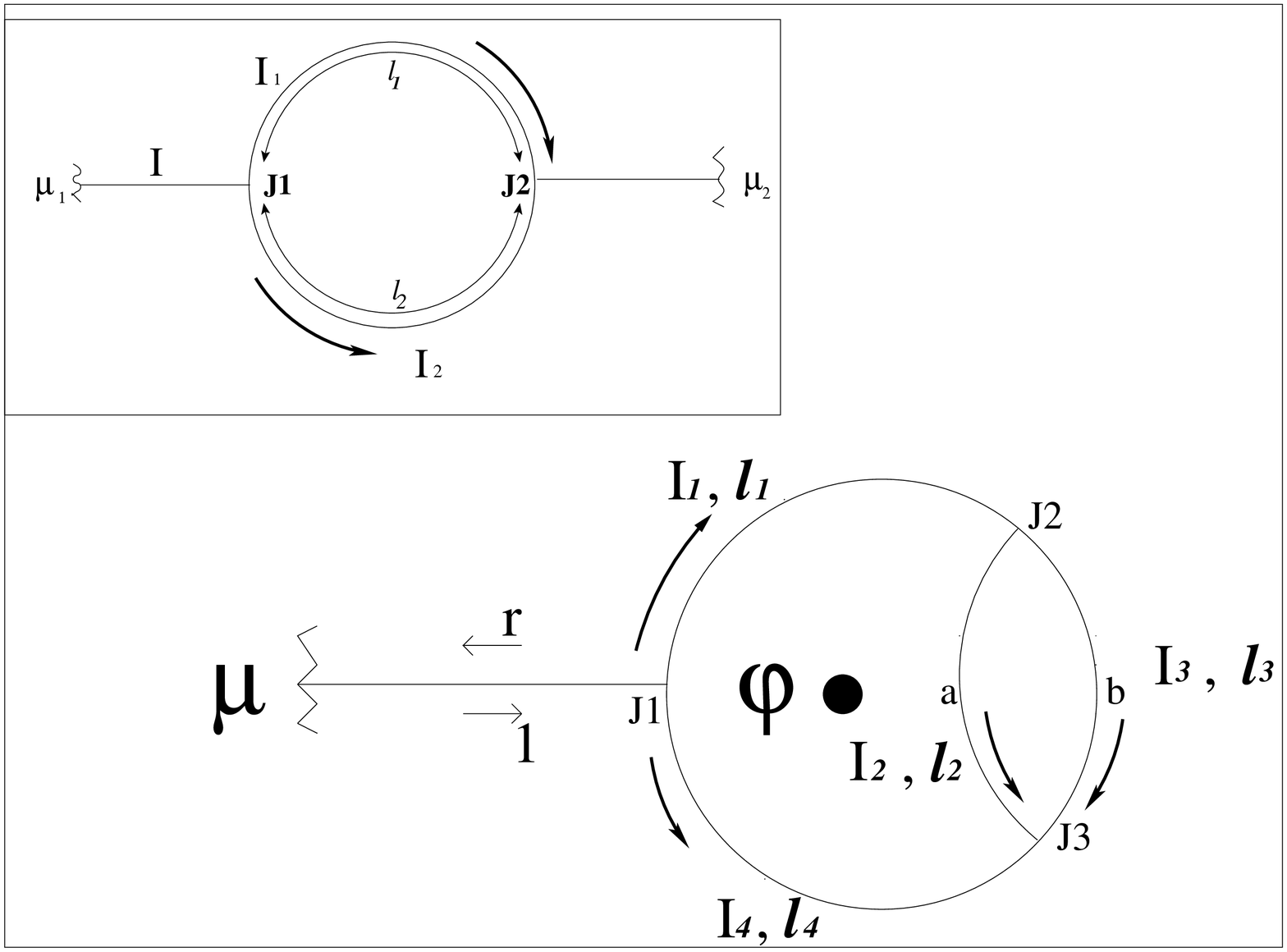}}
\vskip 0.3 in
\caption{The hybrid ring system connected to a reservoir at chemical 
  potential $\mu$. The bubble is denoted by the structure J2bJ3aJ2.The
  localised flux $\Phi$ penetrates the ring. The current densities in
  various parts of the structure are denoted by $I$'s while the lengths
  of the various regions are denoted by $l$'s. In the inset we have
  shown the non-equilibrium case, a one dimensional mesoscopic ring
  with leads is connected to two reservoirs at chemical potentials
  $\mu_{1}$ and $\mu_{2}$.}
\end{figure}

  In the local coordinate system the wave-functions in the various
  regions of our system are given as follows\cite{pohang}-

\begin{eqnarray}
\psi_0&=&  e^{ikx_{0}}+r e^{-ikx_{0}},\nonumber\\
\psi_j&=&a_{j} e^{i(k+\frac{\alpha_{j}}{l_j})x_{j}}+b_{j}
e^{-ikx_{j}+i\frac{\alpha_{j}}{l_j}(x_{j}-l_{j})}.
\end{eqnarray} 

Here $x_{j},j=0,1..4$ are coordinates along the connecting lead to the
reservoir, the segments J1J2, J2aJ3, J2bJ3, and J3J1, while
$\alpha_{j}$'s are the phases picked up by the electron as it
traverses the various regions of the system and $l_j$'s are the
lengths of the segments J1J2, J2aJ3, J2bJ3, and J3J1 respectively.
The wave-vector is defined as $k=\sqrt{2mE/\hbar^2}$. To solve for the
unknown coefficients in eqn.~(1) we use
Griffith\cite{grif,xia,deo_stub} boundary condition at the junctions
$J1, J2$ and $J3$. These boundary conditions (given in eqn. 2 below)
are due to the single-valuedness of wavefunction and conservation of
current (Kirchoff's law).

\begin{eqnarray}
\psi_{0}(x_0=0)=\psi_{1}(x_1=0)=\psi_{4}(x_4=l_{4}),\nonumber\\
\psi_{0}^\prime(x_0=0)+\psi_{4}^\prime(x_4=l_{4})=\psi_{1}^\prime(x_1=0),\nonumber\\
\psi_{1}(x_1=l_{1})=\psi_{2}(x_2=0)=\psi_{3}(x_3=0),\nonumber\\
\psi_{1}^\prime(x_1=l_{1})=\psi_{2}^\prime(x_2=0)+\psi_{3}^\prime(x_3=0),\nonumber\\
\psi_{2}(x_2=l_{2})=\psi_{3}(x_3=l_{3})=\psi_{4}(x_4=0),\nonumber\\
\psi_{2}^\prime(x_2=l_{2})+\psi_{3}^\prime(x_3=l_{3})=\psi_{4}^\prime(x_4=0).
\end{eqnarray} 
Herein $\psi_j^\prime(x_j=l_j)$ denotes $[(\frac{\partial}{\partial x_j}-
\frac{i \alpha_j}{l_j})\psi_j]_{x_j=l_j}$.
Using the above mentioned boundary conditions we get-

\begin{eqnarray}
1 + r = a_{1} + b_{1} e^{-i\alpha_{1}}= a_{4} e^{ikl_{4}+ i\alpha_{4}}
+ b_{4} e^{-ik l_{4}},\nonumber\\
1 - r - a_{1} +b_{1} e^{-i\alpha_{1}}+  a_{4}e^{ikl_{4}+
  i\alpha_{4}}-b_{4} e^ {-ikl_{4}} = 0 ,\nonumber\\  
a_{1} e^{ikl_{1}+ i\alpha_{1}}+ b_{1} e^{i k l_{1}}= a_{2} + b_{2}
e^{i\alpha_{2}}= a_{3} + b_{3} e^{i\alpha{3}},\nonumber\\
a_{1} e^{ikl_{1}+ i\alpha_{1}} - b_{1} e^ {-i k l_{1}} - a_{2} + b_{2}
e^{-i\alpha_{2}} - a_{3} + b_{3} e^{-i\alpha_{3}} = 0,\nonumber\\
a_{2} e^{ikl_{2}+i\alpha_{2}} + b_{2} e^ {-i k l_{2}} = a_{3} e^{ikl_{3}+
i\alpha_{3}}+ b_{3} e^{- ikl_{3}} = a_{4} + b_{4} e^{- i\alpha_{4}},\nonumber\\
a_{2} e^{ikl_{2}+ i\alpha_{2}} - b_{2} e^{-i k l_{2}} + a_{3} e^{ikl_{3}+
i\alpha_{3}} - b_{3} e^{-ikl_{3}} - a_{4} + b_{4} e^{-i\alpha_{4}}= 0. 
\end{eqnarray}

  Here $\alpha_{1},\alpha_{2},\alpha_{3}$ and $\alpha_{4}$ are phases
  picked up by the wave-functions in the segments J1J2, J2aJ3, J2bJ3
  and J3J1 respectively and we have
  $\alpha_{1}+\alpha_{2}+\alpha_{4}=2\pi\Phi/\Phi_{0},$ and
  $\alpha_{1}+\alpha_{3}+\alpha_{4}=2\pi\Phi/\Phi_{0}$ such that
  $\alpha_{2}=\alpha_{3}$ as required by definition. Using eqn. (3) we
  have solved for all the unknown coefficients in eqn. (1).
  
  In the lead connecting the reservoir to our circuit there is no
  current flow as $|r|^2=1$.  The current densities ($I_j$, in a
  dimensionless form)\cite{buti} in the small interval $dk$ around the
  Fermi energy $k$ in the various segments of the circuit are given
  by- $I_{j}=|a_{j}|^2-|b_{j}|^2$. The current densities are
  calculated from the usual formula of current density in presence of
  magnetic flux-

\begin{eqnarray}
J_{j}=\frac{e\hbar}{2mi}(\psi_j^*\nabla\psi_j-
\psi_j\nabla\psi_j^*-\frac{2i\alpha_{j}}{l_j}\psi_j^*\psi_j),
\end{eqnarray}

which implies $I_{j}=\frac{J_{j}}{e\hbar k/m}$.

The persistent current densities in various parts of the circuit show
cyclic variation with flux and $\Phi_{0}$ periodicity, and oscillate
between positive and negative values as a function of energy or the
wavevector $k$ as expected. Since the analytical expressions for these
currents are too lengthy we confine ourselves to a graphical
interpretation of the results. It should be noted that in all these
expressions for current densities flux enters only through the
combinations $\alpha_{1}+\alpha_{2}+\alpha_{4}$ and
$\alpha_{1}+\alpha_{3}+\alpha_{4}$ the magnitude of these combinations
is given by $2\pi\Phi/\Phi_{0}$ as expected. For us the current
densities in the bubble, i.e., J2bJ3aJ2 are of special importance as
in this region there is a possibility of current enhancement which
will be analysed below.  The currents flowing in segment J3J1 and J1J2
are equal, i.e., $I_{1}=I_{4}$. These currents may have positive
(clockwise) or negative (anti-clockwise) values depending on the flux
$\Phi$ and value of wavevector $k$. For a fixed $k$ this current
oscillates between positive and negative values as a function of
$\Phi$ with a period $\Phi_{0}$ and are asymmetric in $\Phi$.
Similarly for fixed value of $\Phi$ currents oscillate as one varies
$k$. The magnitude of current shows a maximum or minimum near the
corresponding eigen-states of the system. We have calculated these
eigen states for two different cases. For open system as depicted in
figure~1 one can calculate the energies (or wave-vector) of these
states by looking at the complex poles of the S-Matrix. In our case
S-Matrix is simply a complex reflection amplitude $r$. We have also
analysed the eigen states of a closed system (without coupling lead to
reservoir) by wave-function matching in various segments using
waveguide theory.
\begin{figure}{h}
\protect\centerline{\epsfxsize=3.0in \epsfbox{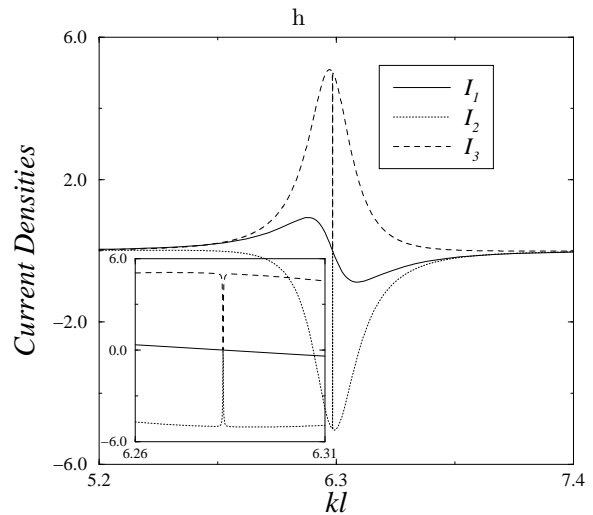}}
\caption{Current enhancement shown with lengths 
  $l_{1}/l=l_{4}/l=0.75,$ $l_{2}/l= 0.45,$ $l_{3}/l=0.55$. Herein the 
  persistent current densities in the various parts of the circuit are
  plotted. The persistent current density in $J1J2$ is denoted by the
  solid line while those in $J2aJ3$ and $J2bJ3$ are denoted by dotted
  and dashed line. Flux $\alpha=2\pi\Phi/\Phi_0$=0.1 .}
\end{figure}

\begin{figure}{h}
\protect\centerline{\epsfxsize=3.0in\epsfbox{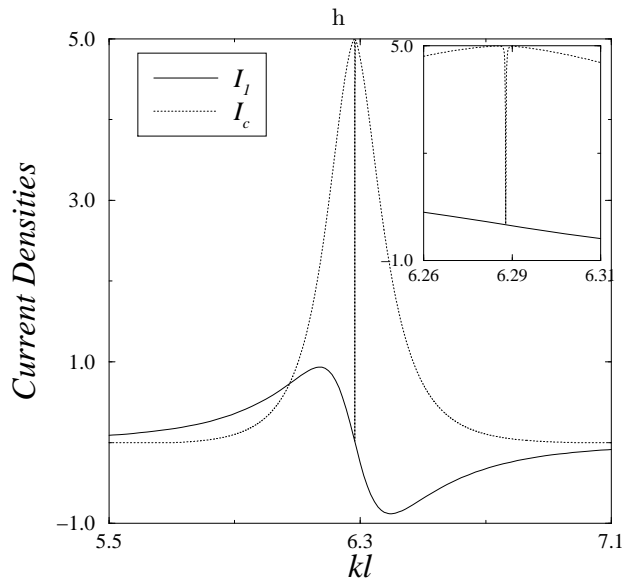}}
\caption{Persistent current density $I_{1}$ and circulating  current
  density $I_{c}$ is plotted as function of $kl$. The parameters are
  same as used in fig.~2. The inset shows the behavior of $I_{c}$ and
  $I_{1}$ around their zero values.}
\end{figure}

We analyse the case of a bubble with unequal lengths, of its two arms,
i.e., the length of $J2bJ3$ $\neq$ $J2aJ3$. This asymmetry implies
that currents in the two arms of the bubble are unequal, i.e., $I_{2}$
$\neq$ $I_{3}$. This asymmetry is very much essential for current
enhancement. In figure~2 we plot the persistent current densities in
various parts of the circuit, for this we have chosen a small interval
of dimensionless wavevector $kl$ and other physical parameters are
indicated in the figure captions.  It should be noted that the
absolute value of the persistent current densities $I_{2}$ and $I_{3}$
in this interval are individually much larger than the input current
density $I_{1}$ into the bubble and thus the current enhancement
effect is evident (without violating the basic Kirchoff's law). The
input current arises due to the presence of flux $\Phi$ as it breaks
the time reversal symmetry. In the interval $5.2<kL<7.4$ current
$I_{1}$ changes from positive to negative and exhibits extremum around
the real part of the poles of the S-Matrix.  When $I_{1}$ is positive,
negative current density of magnitude $I_{2}$ flows in the arm $J2aJ3$
of the bubble.  Thus, when $I_{1}$ is positive circulating current
flows in the clockwise direction in the bubble. In the range where
$I_{1}$ is negative, i.e., input current into the bubble is in an
anti-clockwise direction, then positive current flows in arm $J2bJ3$.
According, to our convention as mentioned earlier, circulating current
flows in the clockwise direction. The magnitude of this circulating
current $I_{c}$ (negative current), is taken to be the value of
current in one of the arms of the bubble moving against the input
current into the bubble as explained in detail in the introduction. In
all the figures drawn the length of the bubble $l=l_{2}+l_{3}$ is
taken as unity, and the current densities along with the Fermi
wave-vectors are in their dimensionless form. This current enhancement
effect is extremely sensitive to the lengths of the arms of the
bubble.  In figure ~3 we have plotted the persistent current density
$I_{1}=I_{4}$ and the circulating current density $I_{c}$ in the
bubble for the same parameters used in figure~2. For figure~3 we have
used a smaller scale of $kl$ for clarity. Even though persistent
current density $I_1$ changes the sign as we cross the real part of
the pole the circulating current remains clockwise only. Moreover,
$I_c$ around either side of the poles has a magnitude larger than the
absolute magnitude of $I_1$.  It should be noted that if we
interchange the values of $l_{2}$ and $l_{3}$ keeping other parameters
unchanged circulating current will flow in an anti-clockwise
direction. This is obvious from the geometry of the problem.
Alongwith the current densities the total persistent currents in
various parts of the ring can also be plotted, to do that we integrate
the current densities $J_{j}$ in various regions of the circuit over
the Fermi wave vector. The persistent currents $P_{j}$ at temperature
$T=0$ is given by

\begin{eqnarray}
P_{j}=-\int_{0}^{k_{f}} J_{j} dk 
\end{eqnarray}

In figure~4 we have plotted the persistent currents (in dimensionless
units) as a function of flux for a fixed value of $k_f$ and the system
parameters are indicated in the figure caption as expected persistent
currents in various arms are flux periodic and asymmetric in flux.
Here also we can see the persistent current enhancement effect in some
range of flux $\Phi$.
\begin{figure}{h}
\protect\centerline{\epsfxsize=3.0in\epsfbox{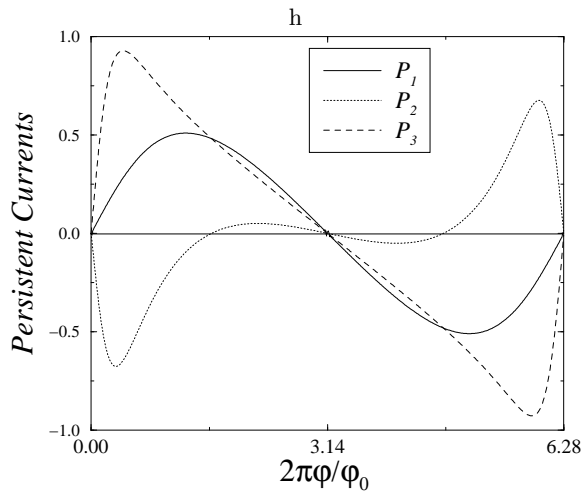}}
\caption{Current enhancement shown with lengths 
  $l_{1}/l=l_{4}/l=0.25,$ $l_{2}/l=0.45,$ $l_{3}/l=0.55$. Herein the
  persistent currents in the various parts of the circuit are plotted
  as function of flux. The Fermi wavevector here is $k_{f}=4\pi$.}
\end{figure}

We generally observe enhanced currents at those Fermi energy
wave-vector intervals which are around the poles of the open
system\cite{deo_cm,pareek_cm,physicapareek}. However, there are some
exceptions. In figures~5 and 6 we plot a few of those exceptions. We
consider the lengths $l_{1}/l=l_{4}/l=0.75, l_{2}/l=0.35$ and
$l_{3}/l=0.65$ and flux $\alpha=\frac{2\pi\Phi}{\Phi_0}=0.1$. In figures~5
and 6 we show that current enhancement effect does not occur at places
which are eigen values of the aforesaid system. Here the eigen
wave-vector $kL$ corresponds to $11.28$ in figure~5 and $13.85$ in
figure~6. One can readily notice that the magnitude of persistent
current density (i.e., input current density $I_{1}$) shows extrema
around this value. Around this region the current densities in the
bubble $I_{2}$ and $I_{3}$ are individually smaller than $I_{1}$ and
they flow in the same direction as the input current.  Hence we do not
observe current enhancement effect around the quasi bound states for
these parameters of the open system. The exact conditions for current
enhancement cannot of course be readily predicted \emph{a priori}.  If
system exhibits current enhancement one should be able to detect it
experimentally by observing the enhanced response of the magnetic
moment.
\begin{figure}{h}
\protect\centerline{\epsfxsize=3.0in \epsfbox{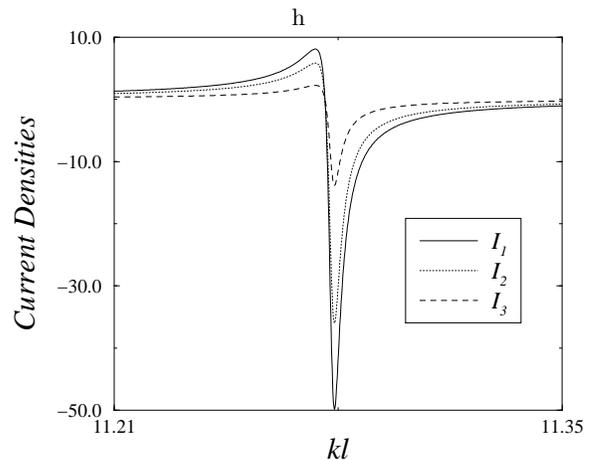}}
\caption{Absence of  current enhancement shown with lengths
  $l_{1}/l=l_{4}/l=0.75,$ $l_{2}/l= 0.25,$ $l_{3}/l=0.75$. Herein the
  persistent current densities in the various parts of the circuit are
  plotted. The persistent current density in $J1J2$ is denoted by the
  solid line while those in $J2aJ3$ and $J2bJ3$ are denoted by dotted
  and dashed line. The $kl$ value $11.28$ is an eigen wavevector of
  the closed system. Flux =0.1 .}
\end{figure}

\begin{figure}{h}
\protect\centerline{\epsfxsize=3.0in \epsfbox{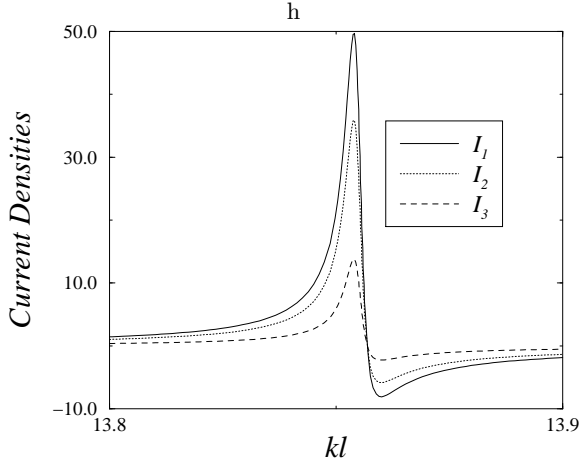}}
\caption{Absence of  current enhancement shown with lengths
  $l_{1}/l=l_{4}/l=0.75,$ $l_{2}/l= 0.25,$ $l_{3}/l=0.75$. Herein the
  persistent current densities in the various parts of the circuit are
  plotted. The persistent current density in $J1J2$ is denoted by the
  solid line while those in $J2aJ3$ and $J2bJ3$ are denoted by dotted
 and dashed line. The $kl$ value $13.85$ is an eigen wavevector of
  the closed system. Flux =0.1 .}
\end{figure}

As mentioned in the paragraphs above, current enhancement directly
corresponds to resonances of our system. These resonances occur at or
near the quasi bound states of the open system. Thus it is important
to study the closed system analog to see where these bound states
occur and how are they linked to the quasi bound states. More than
that the study of bound states of the closed system will give us the
eigen energy spectra from which many important parameters like
magnetic moment and persistent currents can be evaluated.  In
figures~7,8 and 9 we have plotted the first few eigen energies
$E=k_{n}^2$ of our isolated system (with the connecting lead to
reservoir removed). The parameters are mentioned in the figure
captions. These eigen energies are calculated with the help of quantum
waveguide approach from the condition that the determinant of the
coefficient matrix must vanish.  The coefficient matrix is built from
first principles using quantum waveguide theory with the second
wavefunction of equation~1. The condition for bound states of the
system, i.e., the closed ring with bubble is given by

\begin{eqnarray}
 \cos(\alpha) & = & \frac{1}{\cos(kl_{-})} (\cos k(l_{1}+l_{+}) - \nonumber\\
              &   &\frac{1}{4} \frac{\sin (kl_{1}) \sin (kl_{2})
                   \sin (kl_{3})}{\sin (kl_{+})})
\end{eqnarray}

where $\alpha=2\pi\Phi/\Phi_{0}$, $l_{+}=(l_{2}+l_{3})/2$ and
$l_{-}=(l_{2}-l_{3})/2$. The eigen energies are flux periodic. The
persistent current carried by a electron in the eigen state $E_{n}$ is
given by $I_{n}=-\frac{1}{c}\frac{\partial E_{n}}{\partial\Phi}$. In a
closed single loop persistent current changes its sign as we go from
one level to the next successive level. Thus for spinless electrons
the persistent current is diamagnetic or paramagnetic depending on
total number electrons being odd or even(at $T=0$).  This is called
parity effect. Supposing we open up the ring and find its transmission
amplitude $t$ then we can easily see that

\begin{eqnarray}
 \cos(\alpha) & = & Re(1/t)
\end{eqnarray}

In our system with the help of eqn.~7 we can calculate the persistent
current carried by an electron in the eigen state $E_n$. This is given
by- $I_{n}=-\frac{e}{\hbar}\frac{sin(\alpha)}{{d Re(1/t)/d E_n}}$.
From the above formula we get for the persistent currents in our
closed system the following expression-

\begin{eqnarray}
\frac{\hbar I_{n}}{e}=\frac{-8k_{n}sin(\alpha)cos^2(k_{n}l_{-})
 sin^2(k_{n}l_{+})}{D}
\end{eqnarray} 
where
$D=sin(k_{n}l_{3+}(16l_{2+}+2l_{3+})+sin(k_{n}l_{2+}(16l_{3+}+2l_{2+})
+sin(k_{n}l_{3-})(-6l_{3}-10l_{1}+2l_{3-}cos(2k_{n}l_{2}))
+sin(k_{n}l_{2-})(-6l_{2}-10l_{1}+2l_{2-}cos(2k_{n}l_{3}))
-l_{3+}(sin(k_{n}l_{3+}-2k_{n}l_{2})+9sin(k_{n}l_{3+}+2k_{n}l_{2})
-l_{2+}(sin(k_{n}l_{2+}-2k_{n}l_{3})+9sin(k_{n}l_{2+}+2k_{n}l_{3}))$
and $l_{i\pm}=(l_{1} \pm l_{i})$, herein $i=2,3$ and $k_n $ are the
eigen wave vectors.

\begin{figure}{h}
\protect\centerline{\epsfxsize=3.0in \epsfbox{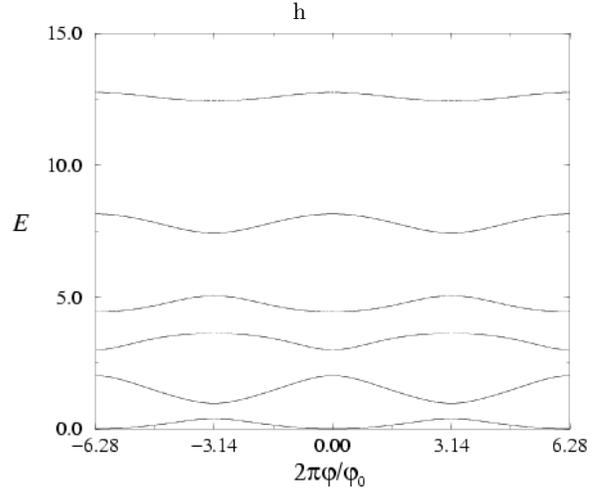}}
\caption{Breakdown of parity effects in a closed one dimensional
  mesoscopic ring coupled to a bubble. The lengths are
  $l_{1}/l=0.75$, $l_{2}/l= 0.35$, $l_{3}/l=0.65$. The energies are
  normalised by $\pi^{2}$.}
\end{figure}

\begin{figure}{h}
\protect\centerline{\epsfxsize=3.0in \epsfbox{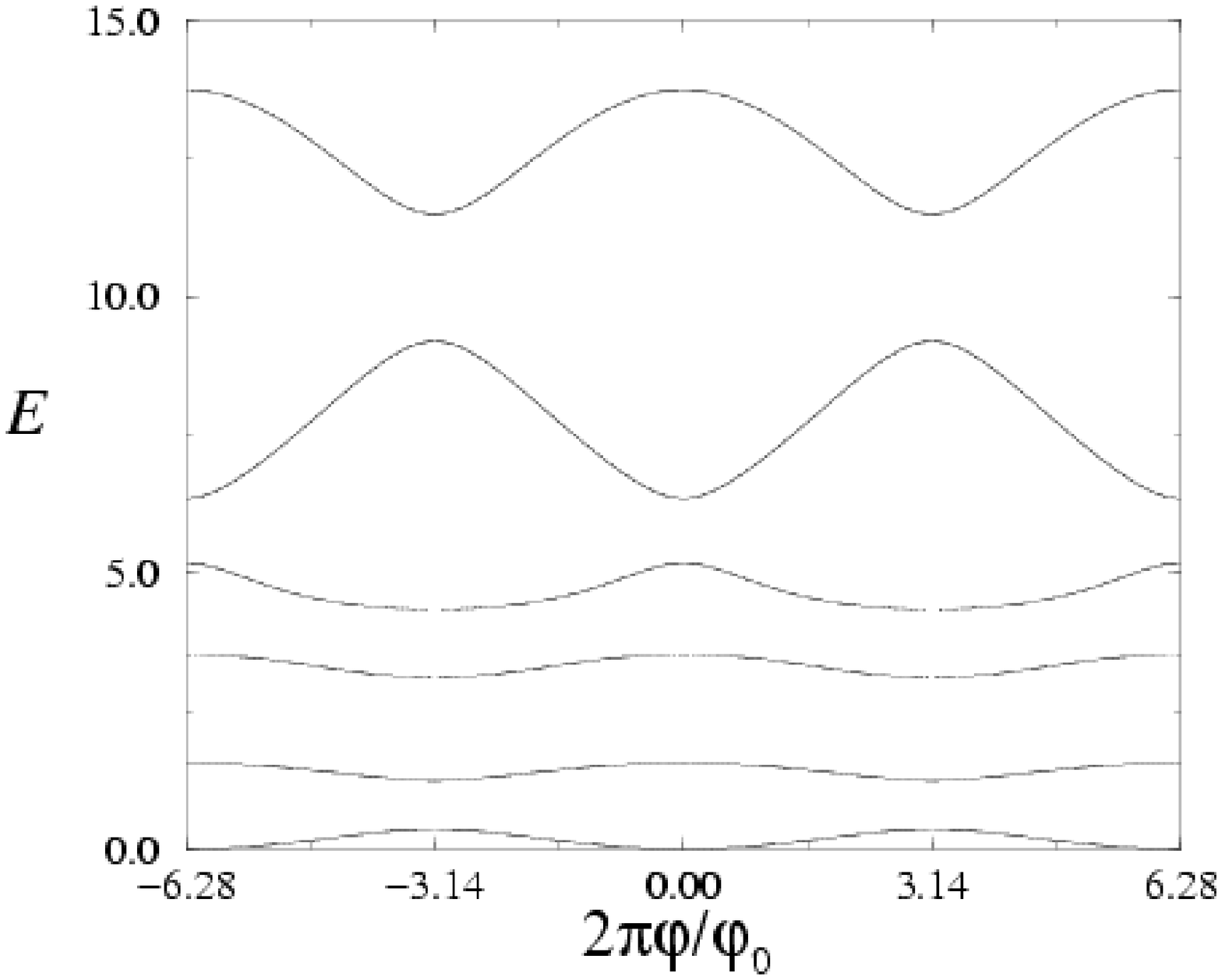}}
\caption{Breakdown of parity effects in a closed one dimensional
  mesoscopic ring coupled to a bubble. The lengths are
  $l_{1}/l=0.75$,$l_{2}/l= 0.15,l_{3}/l=0.85$. The energies are
  normalised by $\pi^{2}$.}
\end{figure}

\begin{figure}{h}
\protect\centerline{\epsfxsize=3.0in \epsfbox{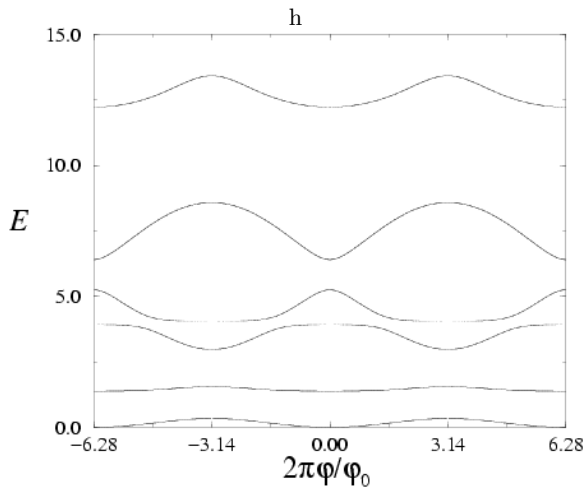}}
\caption{Breakdown of parity effects in a closed one dimensional
  mesoscopic ring coupled to a bubble. The lengths are
  $l_{1}/l=0.75$,$l_{2}/l= 0.05,l_{3}/l=0.95$. The energies are
  normalised by $\pi^{2}$.}
\end{figure}

With the help of $k_n$ one can easily evaluate the persistent currents
from eqn.~8 for some representative values as in figures~7, 8 or 9.
The eigen energies are flux periodic with period $\Phi_{0}$. It can be
noted from figure~7 that 3rd and 4th eigen states carry diamagnetic
current while 5th and 6th carry paramagnetic current for small values
of flux (which is obvious from their slopes). Thus breaking the well
known parity effect. For details we refer to
Ref.[\onlinecite{deo_bpe}]. Similarly, in figures~8 and 9 this
violation of parity effects is also seen but of course the eigen
spectra is modified, although the length of the bubble $l=l_2+l_3=1$
still holds as well as the fact that the length of the outer arm is
also same as in figure~7, what have been altered are the arm lengths
of the bubble, within the restriction that the length of the bubble is
unity. This goes on to show that the eigen energies are sensitive to
the arm lengths of the bubble. We have later on tabulated some of the
eigen $k$ vectors for all these three cases as in the three figures.

\begin{figure}{h}
\protect\centerline{\epsfxsize=3.0in \epsfbox{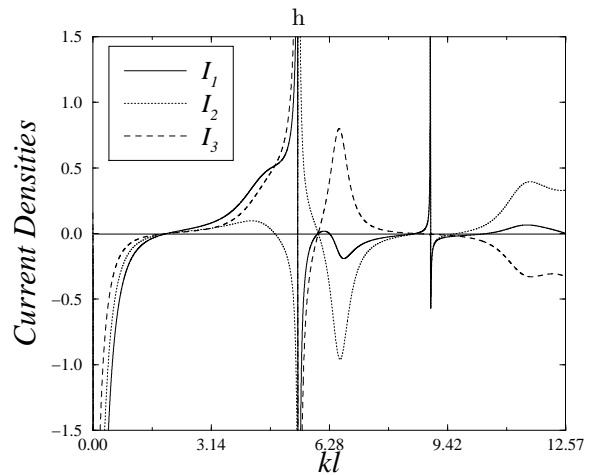}}
\caption{Current densities in the open system for length parameters as 
  in figure~7. Flux = 0.1 .}
\end{figure}

\begin{figure}{h}
\protect\centerline{\epsfxsize=3.0in\epsfbox{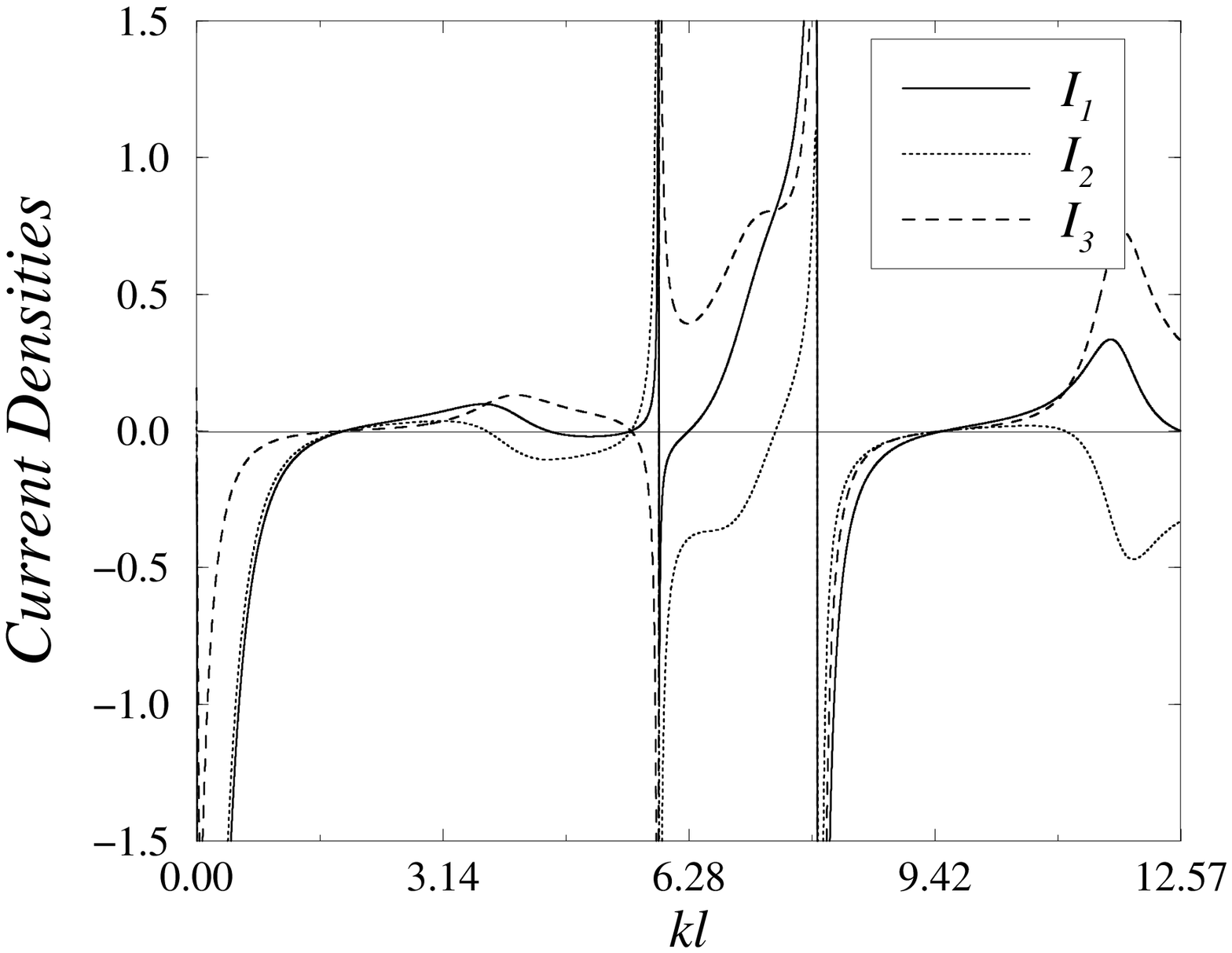}}
\caption{Current densities in the open system for length parameters as 
  in figure~8. Flux = 0.1 .}
\end{figure}

\begin{figure}{h}
\protect\centerline{\epsfxsize=3.0in\epsfbox{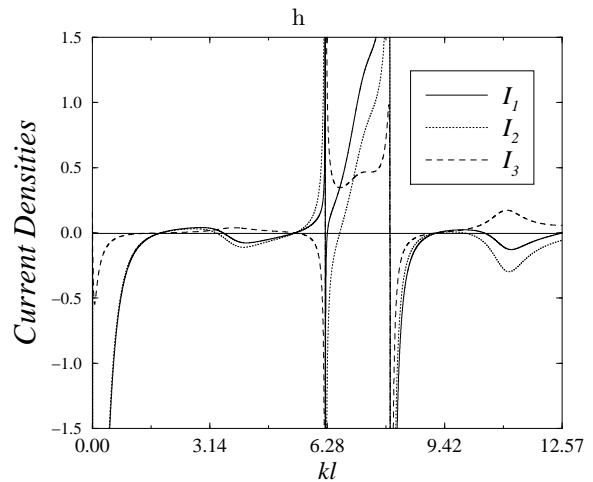}}
\caption{Current densities in the open system for length parameters as 
  in figure~9.  Flux = 0.1 .}
\end{figure}
A comparison of the eigenenergy spectra and the persistent currents in
the open system is herein called for so as to assess for ourselves
whether the persistent currents (global) calculated from the
eigenenergy spectra and inherently linked to the slope of the energy
versus flux diagram are same as the the persistent currents (local) in
the various arms of the system calculated from the eqn.~(3).
Figures~10, 11 and 12 depict the persistent current densities in the
open system for the same parameters as in figures~7, 8 and 9
respectively. These persistent current densities are continuous
functions of the dimensionless wavevector $kl$. We have chosen a
smaller scale for the magnitude of the current densities for better
visibility of the detailed features.  An important conclusion which
can be drawn from the comparisons is that resonances, in the
persistent current densities versus energy graph, arise at the bound
states of the closed system of course they are shifted a bit as a
result of coupling to an external reservoir. The shifted $k$ values
are also later on tabulated.  Another important fact to be noted is
that the energies wherein current enhancement occurs at these, the
local current densities can be either diamagnetic or paramagnetic and
therefore are not same as the global persistent currents but at those
energies wherein current enhancement is not observed therein it is
observed that there is no discrepancy between local and global
currents, i.e., if the eigenenergy spectra reveals diamagnetic
currents the local currents are also seen to be diamagnetic. For
example if we compare figures~7 and 10 we note that the closed system
eigen $k$'s are at 0.1526, 4.4535, 5.4574, 6.627, 8.912, and 11.234.
Around these eigen $k$'s we see from figure~10 that those $k$ values
where current enhancement does not take place therein there is no
difference between the nature (being diamagnetic or paramagnetic) of
global and local current densities in different segments but wherein
current enhancement occurs therein there is a difference between the
nature of global and local current densities (by definition current
enhancement implies this, as the current densities in the two arms of
the bubble are in opposite direction).

\begin{figure}{h}
\protect\centerline{\epsfxsize=3.0in \epsfbox{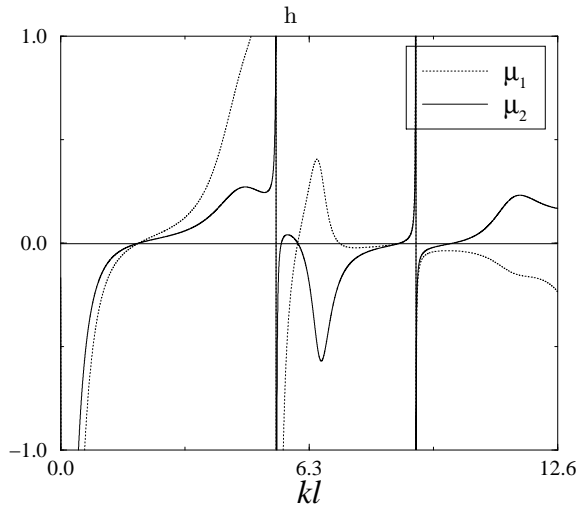}}
\caption{Plot of $\mu_1$ and $\mu_2=I_g A_r/2$ for length parameters
  as in figures~7 and 10. Flux = 0.1 .}
\end{figure}

\begin{figure}{h}
\protect\centerline{\epsfxsize=3.0in \epsfbox{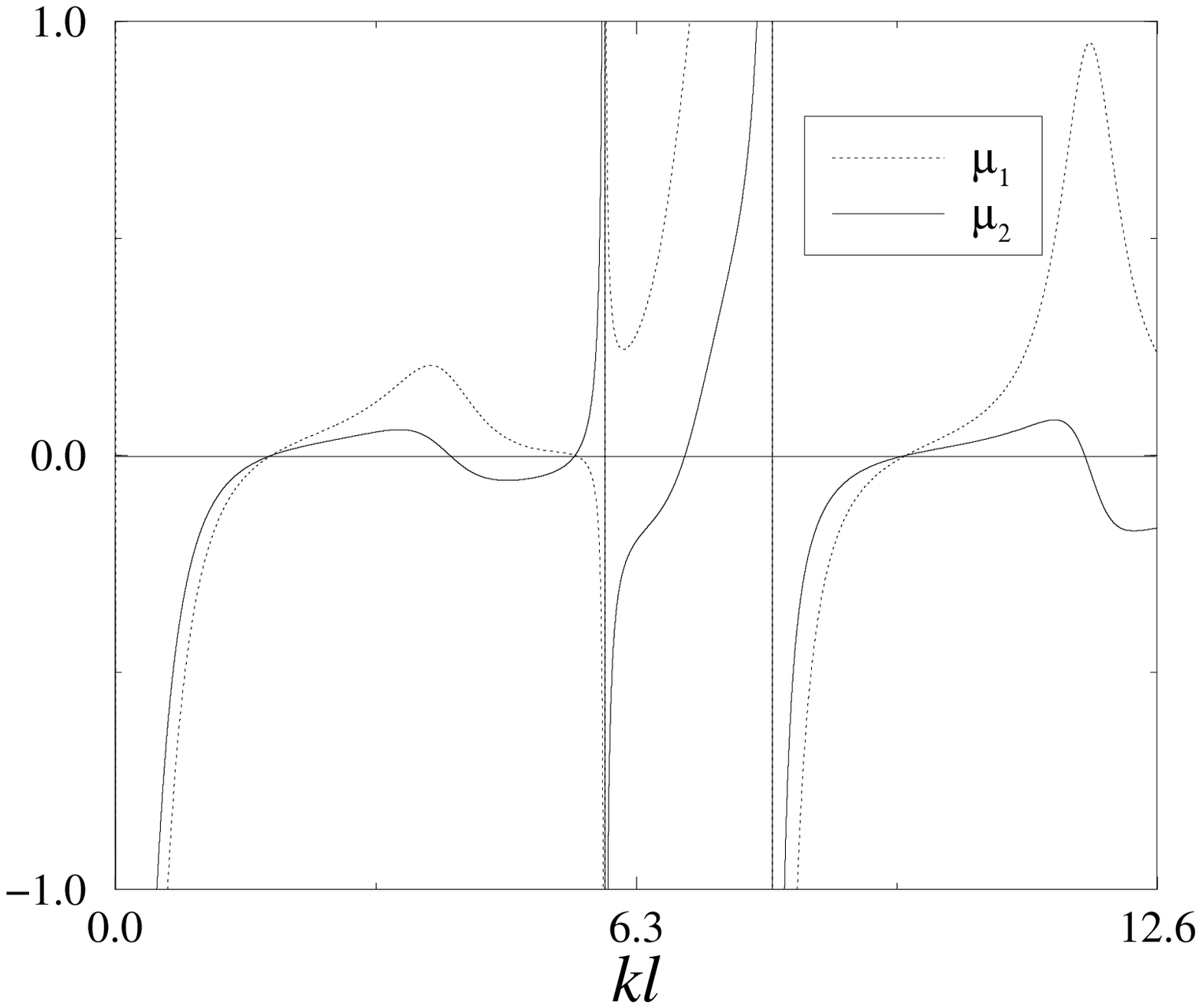}}
\caption{Plot of $\mu_1$ and $\mu_2=I_g A_r/2$ for length parameters
  as in figures 8 and 11. Flux = 0.1 .}
\end{figure}

\begin{figure}{h}
\protect\centerline{\epsfxsize=3.0in \epsfbox{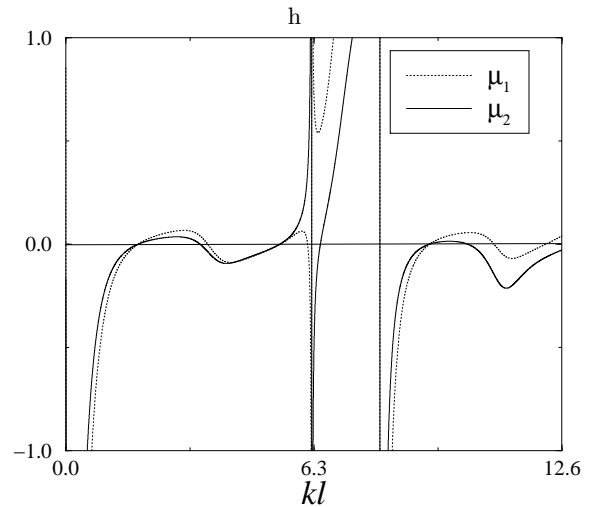}}
\caption{Plot of $\mu_1$ and $\mu_2=I_g A_r/2$ for length parameters
  as in figures 9 and 12. Flux = 0.1 .}
\end{figure}
Extending our discussion to the magnetic moment densities of our
system we see that the magnetic moment densities calculated for the
open system do not seem to agree at least qualitatively with that
calculated from the eigen energy spectra in some energy interval.
Furthermore the orbital magnetic moment density defined via currents
in a loop, depends strongly on the topology of the system, whereas
eigen spectrum do not. In fact there are infinitely many topological
structures possible. If we consider our system as depicted in figure~1
to be planar and lying in the x-y plane then the magnetic moment
density ($\mu_1$) can be viewed as being generated by current density
$I_1$ enclosing an area $A_r$ and by current density $I_3$ enclosing
area $A_b$, i.e., $\mu_1=\frac{1}{c}(I_{1}A_{r}+I_{3}A_{b})$, wherein
$A_{r}$ and $A_{b}$ are areas enclosed by the ring ($J1J2aJ3J1$) and
bubble($J2aJ3bJ2$) respectively. Another orientation of the system in
which the arm $J2bJ3$ is in the x-z plane gives
$\mu_2=\frac{1}{c}(I_{1}A_{r}+I_{2}A_{r})/2=I_g A_r/2$ wherein
$I_{g}=I_{1}+I_{2}$ is said to be the most appropriate generalization
of the equilibrium persistent current, see [\onlinecite{cedraschi}]
for further details, and this along with $\mu_1$ is what we plot in
figures~13, 14 and 15. In these figures, the plot of $\mu_1$ has been
magnified a hundred times for better visibility.  The physical
parameters in these figures, i.e., 13, 14 and 15 are same as that in
figures~7, 8 and 9 respectively (also figures~10, 11 and 12
respectively). Of course $I_g$ is proportional to $\mu_2$. Several
other orientations also are possible, for example, if the bubble lies
in x-y plane and the ring lies in x-z plane, then
$\mu_{z}=\frac{1}{c}(I_{3}A_{r}-I_{2}A_{r})/2$ and
$\mu_{y}=\frac{1}{c}I_{1}A_{r}$. Even when our system lies in the x-y
plane for fixed $l_1, l_2, l_3$ and by deforming their shapes we can
have different values of magnetic moment density along the z
direction. The energy eigen values remain intact inspite of the
deformation. All the above examples buttress the fact that the orbital
magnetic moment is inherently linked to the topology of the system. As
we have earlier pointed out the magnetic moment calculated herein is
not same (qualitatively), as the magnetic moment calculated in case of
the closed system from its eigenenergy spectra, which is same for all
topological situations. Some important conclusions can be drawn from
the figures~13, 14 and 15, first $\mu_1$ and $\mu_2$ are not same and
also one can notice that in some $kl$ intervals $\mu_1$ and $\mu_2$
have opposite signs, i.e., the nature of magnetic moment is
qualitatively different for these two cases. Thus this leads to a
contradiction as to with which of these can one associate the magnetic
moment. The magnetic moment is fixed for closed system from the energy
eigenvalue spectrum calculated from $\mu=-\frac{1}{c}\frac{\partial
  E_n}{\partial H}$ wherein $H$ is the magnetic field enclosed by the
system.  In the open system we expect behavior as a function of $kl$
to reflect the behavior in the closed system except for the fact that
we get currents for all the $kl$ wave vectors. This arises from the
fact that in open system life time of a particle in an eigen state is
finite due to coupling to the reservoir and hence all the energy
eigenvalues associated with the closed system get broadened. Again a
comparison between the eigenenergy spectra, and hence magnetic moment
drawn in figures~7,8 and 9 and the figures for magnetic moment
densities (see, figures~13, 14 and 15) shows that in some cases the
range wherein successive energy eigenvalue slopes are similar (i.e.,
either paramagnetic/diamagnetic) the magnetic moment densities
(calculated from currents, e.g., $\mu_1$ or $\mu_2$) can change their
nature. Some cases wherein such changes occur have been tabulated in
the tables~1, 2 and 3. {\it In several cases we notice that magnetic
  moment densities $\mu_1$ or $\mu_2$ have different nature
  (directions) as compared to $\mu$ which is obtained from the slope
  of the energy eigenvalue spectrum. We would have expected the same
  for the magnetic moments calculated from the currents in the
  isolated system}. The coupling to external reservoir does not change
the nature of the currents although it broadens and shifts a bit the
energy eigenvalue spectrum. Thus, the study of magnetic moment
densities of the open system is enough to convey to us the picture
about the magnetic moment calculated from currents in the isolated
system. Hence clarifying our contention.  Thus it is clear that the
nature of magnetic moment densities $\mu_1$ and $\mu_2$ need not be
related to eigenvalue spectrum in the same energy range or equilibrium
magnetic moments at temperature $T=0$. This result we attribute solely
to the current enhancement effect which is purely a quantum mechanical
effect. In the ranges wherein there is no current enhancement, the
nature of magnetic moment densities obtained by two different methods
are qualitatively same, however they may differ in magnitude. It is
also worth mentioning that the total magnetic moment (at temperature
$T=0$) of a representative system is obtained by integrating the
magnetic moment densities upto the Fermi wavevector $k_f$.

The preceding three paragraphs have dealt with the eigen energy
spectra, persistent current densities and the magnetic moments. To
have some conclusive arrivals, we juxtapose all of these in the tables
~1, 2 and 3. In the tables we give the detailed analysis of the eigen
$k$ of the closed system, the poles of the open system, whether at
these poles or bound states current enhancement takes place, what are
the directions (nature) of magnetic moment densities in the open
system in the small interval in the neighborhood of the quasi bound
states, and the persistent currents in the closed system form
calculated from eqn.~8, whose nature is similar to that of the closed
system magnetic moments. In column (1) we give the value of the closed
system eigen $k$, in column (2) we give the complex poles of the open
system for the same lengths. The real part of the pole gives the
energy at which resonance occurs while the imaginary part gives the
width of the same, which of course is just the inverse of the lifetime
of the particle in that resonant state. Thus smaller the imaginary
part, sharper the resonance and more the particles lifetime. In column
(3) we note whether enhanced currents are seen at these quasi-bound
states or not.  In columns (4) and (5) we mention the values of
$\mu_1$ calculated from $I_{1}A_{r}+I_{3}A_{b}$ and $I_g$ in the small
neighborhood of the quasi bound states. In columns for $\mu_1$ and
$I_g$, para/dia implies that around that quasi bound state, nature of
magnetic moment changes from being paramagnetic to the left to being
diamagnetic to the right in the immediate neighborhood of this quasi
bound state (in the same spirit one can understand dia/para in this
column). The areas are calculated by deforming the circular arms of
our system to form rectangles. Finally in column (6) we note the
direction of the magnetic moment of the closed system whose nature
(direction) is same as persistent currents in the closed system as
calculated from eqn.~8.
\begingroup
\squeezetable
\begin{table}[h]
\begin{center}
\begin{tabular}{|c|c|c|c|c|c|c|}
\hline
   $k  $& $poles$& $is$ $current$&$\mu_1$&$I_g$&$\mu$ $from$   \\
   $(1)$& $(2)  $& $enhanced(3) $&$ (4) $&$(5)$&$eigenenergies(6)$\\ \hline
   $0.1526$ &$0.7192-0.044i$&no & dia   & dia   &dia\\ \hline
   $4.4535$ &$4.5157-0.883i$  &no & para  &para   &para\\ \hline
   $5.4574$ &$5.4547-0.005i$  &yes&para/dia&para/dia&dia \\ \hline
   $6.6270$ &$6.5448-0.292i$ &yes& para  & dia   &dia \\ \hline
   $8.9728$ &$8.9728-0.0004i$&yes&para/dia&para/dia&para \\ \hline
   $11.234$ &$11.352-0.748i$  &yes& dia   & para  &para\\ \hline
\end{tabular}
\vskip 0.25in
\caption{\small{A comparative analysis of eigen-energy spectra,
    current densities and magnetic moment density for
    $l_{1}/l=l_{4}/l=0.375,l2/l=0.35,l3/l=0.65$, Refer to figures~7, 10 and 13.
}}\vspace{-6mm}
\end{center}
\end{table}
\endgroup
\begingroup
\squeezetable
\begin{table}[h]
\begin{center}
\begin{tabular}{|c|c|c|c|c|c|c|}
\hline
   $k  $& $poles$& $is$ $current$&$\mu_1$&$I_g$&$\mu$ $from$   \\ 
   $(1)$& $(2)  $& $enhanced(3) $&$ (4) $&$(5)$&$eigenenergies(6)$\\ \hline
   $0.160$&$0.162-0.571i $&no & dia   & dia    &dia\\ \hline
   $3.921$&$3.957-0.670i $&yes & para &para    &para\\\hline
   $5.898$&$5.898-0.001i $&yes&dia/para&para/dia&para\\\hline
   $7.136$&$7.084-0.667i $& no &para  & para   &para\\ \hline
   $7.948$&$7.940-0.012i $& no & dia   & dia    & dia\\ \hline
   $11.65$&$11.754-0.438i$& yes& para & para   &para\\ \hline
\end{tabular}
\vskip 0.1in
\caption{\small{A comparative analysis of eigen-energy spectra,
    current densities and magnetic moment density for
    $l_{1}/l=l_{4}/l=0.375$, $l_2/l=0.15,$ $l_3/l=0.85$. Refer to
    figures~8, 11 and 14.  }}\vspace{-5mm}
\end{center}
\end{table}
\endgroup
\begingroup
\squeezetable
\begin{table}[h]
\begin{center}
\begin{tabular}{|c|c|c|c|c|c|c|}
\hline
   $k  $& $poles$& $is$ $current$&$\mu_1$&$I_g$&$\mu$ $from$   \\ 
   $(1)$& $(2)  $& $enhanced(3) $&$ (4) $&$(5)$&$eigenenergies(6)$\\ \hline
   $0.169$&$0.183-0.5660i$ & no & dia   & dia   &dia\\ \hline
   $3.693$&$3.706-0.6380i$ &yes & para  & para  &dia\\ \hline
   $6.228$&$6.228-0.0004i$ &yes &dia/para&para/dia&para\\ \hline
   $7.170$&$7.160-0.6530i$ & no &para   & para  &para\\ \hline
   $7.989$&$7.981-0.0154i$ & no & dia   & dia   & dia\\ \hline
   $11.00$&$11.019-0.551i$ & yes& dia   &dia    &dia\\ \hline
\end{tabular}
\vskip 0.25in
\caption{\small{A comparative analysis of eigen-energy spectra,
    current densities and magnetic moment density for
    $l_{1}/l=l_{4}/l=0.375,$ $l_2/l=0.05$, $l_3/l=0.95$. Refer to
    figures~9, 12 and 15.  }}\vspace{-5mm}
\end{center}
\end{table}
\endgroup

From the tables and graphs, one can draw some conclusions:

(1). At eigen-energies where there is no current enhancement, the
entries of column (5) are compatible with that of column (6) but at
places where current enhancement occurs the entries may or may not
match.

(2). Apart from the above inconsistency, we also observe from the
graphs for eigenenergy spectra and magnetic moment and current
densities that for parameters at which over successive energy levels,
persistent currents are paramagnetic (for example, see figure~(8)),
the current (figure~(11)) and magnetic moment densities (figure~(14))
are not paramagnetic, i.e., in between energy levels they do change
sign.

Hence, the parameter $I_g$ hitherto alluded to as the the most
effective generalisation of the equilibrium persistent current is not
necessarily correct. In fact we cannot think of any other topological
configuration for which magnetic moments calculated from the current
densities and that from the eigenvalue spectra (equilibrium
magnetization) would match.

To conclude, we have shown that current enhancement effect can occur
in equilibrium mesoscopic systems in presence of magnetic flux.
Earlier, it was shown that this effect arises in a non-equilibrium
state, i.e., in presence of transport current flow and in absence of
magnetic field. This quantum effect is extremely sensitive to system
parameters. Parity effects are shown to be violated in the isolated
system. Apart from this we have analysed the global and local current
densities of our system and shown that the orbital magnetic response
of the system calculated from the current densities (and inherently
linked to the topological configuration) is qualitatively not same as
that calculated from the eigenenergy spectra. This fact is related to
the current enhancement effect in these systems. To clarify our
preceding contention it is imperative for us to study a system which
does not exhibit current magnification effect. A suitable system to
study these local and global currents and the differences which occur
between them would be a system of coupled rings. Herein two different
rings are coupled via an ideal lead as studied in Ref.
[\onlinecite{couple}]. In this system there is no question of current
enhancement. However, currents (or magnetic moments) in the two rings
may be of opposite sign, in this case one can also obtain the global
currents (or magnetic moments) from the eigen energy spectra. It is
worthwhile to check whether the sum of magnetic moments obtained from
currents of two separate rings shows the same nature as obtained from
the eigen energies of the entire system.

\acknowledgments
Authors acknowledge help of, and useful discussions with Sandeep K. Joshi.

\end{document}